# Vanadium oxide and a sharp onset of cold-trapping on a giant exoplanet

Stefan Pelletier[1]†, Björn Benneke[1], Mohamad Ali-Dib[2], Bibiana Prinoth[3], David Kasper[4], Andreas Seifahrt[4], Jacob L. Bean[4], Florian Debras[5], Baptiste Klein[6], Luc Bazinet[1], H. Jens Hoeijmakers[3], Aurora Y. Kesseli[7], Olivia Lim[1], Andres Carmona[8], Lorenzo Pino[9], Núria Casasayas-Barris[10], Thea Hood[5], and Julian Stürmer[11]

[1] Department of Physics and Trottier Institute for Research on Exoplanets, Université de Montréal, Montreal, QC, Canada
[2] Center for Astro, Particle and Planetary Physics, New York University Abu Dhabi, UAE
[3] Lund Observatory, Division of Astrophysics, Department of Physics, Lund University, Lund, Sweden
[4] Department of Astronomy & Astrophysics, University of Chicago, Chicago, IL, USA
[5] Université de Toulouse, CNRS, IRAP, Toulouse, France
[6] Department of Physics, University of Oxford, Oxford, UK
[7] IPAC, Caltech, Pasedena, CA, USA
[8] Université Grenoble Alpes, CNRS, IPAG, Grenoble, France
[9] INAF-Osservatorio Astrofisico di Arcetri Largo Enrico Fermi, Florence, Italy
[10] Leiden Observatory, Leiden University, Leiden, The Netherlands
[11] ZAH Landessternwarte, Heidelberg, Germany
† email: stefan.pelletier@umontreal.ca

**The abundance of refractory elements in giant planets can provide key insights into their formation histories[1]. Due to the Solar System giants' low temperatures, refractory elements condense below the cloud deck limiting sensing capabilities to only highly volatile elements[2]. Recently, ultra-hot giant exoplanets have allowed for some refractory elements to be measured showing abundances broadly consistent with the solar nebula with titanium likely condensed out of the photosphere[3,4]. Here we report precise abundance constraints of 14 major refractory elements on the ultra-hot giant planet WASP-76b that show distinct deviations from proto-solar, and a sharp onset in condensation temperature. In particular, we find nickel to be enriched, a possible sign of the accretion of a differentiated object's core during the planet's evolution. Elements with condensation temperatures below 1,550 K otherwise closely match those of the Sun[5] before sharply transitioning to being strongly depleted above 1,550 K, well explained by nightside cold-trapping. We further unambiguously detect vanadium oxide on WASP-76b, a molecule long hypothesized to drive atmospheric thermal inversions[6], and also observe a global east-west asymmetry[7] in its absorption signals. Overall, our findings indicate that giant planets have a mostly stellar-like refractory elemental content and suggest that temperature sequences of hot Jupiter spectra can show abrupt transitions wherein a mineral species is either present, or completely absent if a cold-trap exists below its condensation temperature[8].**

We observed three transits of the ultra-hot Jupiter WASP-76b[9] using the M-dwarf Advanced Radial velocity Observer Of Neighboring eXoplanets (MAROON-X) high-resolution optical spectrograph[10] at the 8.1-m Gemini-North Observatory in Hawaii. With a continuous wavelength coverage between 490 and 920 nm at a spectral resolution of $\lambda/\Delta\lambda \approx 85,000$, MAROON-X is well suited for probing atomic species in planetary atmospheres. Two transits of WASP-76b were observed on 2020 September 3 and 12, and a third transit on 2021 October 28. Each transit observation consists of a time series of between 36 and 47 high-resolution spectra, each with integration times between five and eight minutes.

The observed spectra contain contributions from WASP-76b (the planet), WASP-76 (the host star), and from the Earth's atmosphere. The latter two dominate the spectra and must be removed in order to uncover the comparatively much fainter planetary signal. We achieve this by employing a Principal Component Analysis (PCA) based algorithm which removes the relatively stationary-in-velocity stellar (<1 km s$^{-1}$) and telluric (0 km s$^{-1}$) contributions from the data while leaving the rapidly Doppler shifting (~100 km s$^{-1}$) planetary signal largely unaffected (see Methods and Extended Data Fig. 1). To uncover WASP-76b's atmospheric signature, we cross-correlate the resulting cleaned-up spectra with transit model templates as a function of radial velocity. If a model matches the data well, the cross-correlation function (CCF) of each in-transit spectrum should peak at the exact Doppler shift matching WASP-76b's orbital velocity at that moment[11]. Viewing the cross-correlation as a function of time should then produce a trail that is centered at the velocity of the system ($V_{sys}$) with a slope matching WASP-76b's radial velocity semi-amplitude ($K_p$). The CCFs at each phase can then be summed for different configurations of $K_p$ and $V_{sys}$ to produce two-dimensional signal-to-noise maps for a given atmospheric model. If a species is detected in WASP-76b's atmosphere, a strong peak will be observed in the resulting map near the expected $K_p$ and $V_{sys}$. We repeat this cross-correlation process using model templates containing absorption lines of only a single metal, ion, or molecule (Extended Data Fig. 2) at a time to produce a chemical inventory of WASP-76b's atmosphere.

We detect Fe, Na, Ca$^+$, Cr, Li, H, V, VO, Mn, Ni, Mg, Ca, K, and Ba$^+$ in WASP-76b's atmosphere, as well as tentatively detect O and Fe$^+$, providing an unprecedentedly complete measurement of the chemical inventory of a gas giant planet (Fig. 1). We also notably do not find evidence of absorption from neutral or oxidized titanium (Extended Data Fig. 3), despite being highly sensitive to those species if they were in chemical equilibrium abundances. The observed presence or absence of these species provides us with key insight into the conditions of WASP-76b's highly irradiated atmosphere. Vanadium oxide (VO), in particular, is a strong optical absorber that has long been sought in ultra-hot Jupiter atmospheres due to its theorized role as a driver of thermal inversions[6]. We detect the VO signal in all three individual MAROON-X transits, and also with an independent analysis of two ESPRESSO transits of WASP-76b (see Methods and Extended Data Fig. 4). Our detection thus confirms that VO is present in hot Jupiter atmospheres as a source of atmospheric heating, alongside other shortwave absorbers such as H$^-$ and atomic metals[12]. Especially in the absence of titanium oxide, another highly potent optical absorber present in some even hotter exoplanets[13], the VO molecules are directly exposed to the incoming short-wavelength stellar irradiation and become the dominant optical broadband absorbers above the H$^-$ continuum, adding hundreds of Kelvins to WASP-76b's upper atmosphere[14]. The detection of ionized barium, with an atomic number $Z = 56$, also shows that, like Jupiter[15], WASP-76b's atmosphere is not significantly fractionated by mass. Additionally, the combined presence of V/VO, Ca/Ca$^+$, and Fe/Fe$^+$ provides a physical and chemical thermometer wherein WASP-76b's atmosphere must have temperature regions where these combinations of species can coexist, albeit potentially at different altitudes and/or longitudes. Most detections are slightly offset in velocity space from literature predictions for a symmetric and static atmosphere, indicative of dynamical[7] and chemical inhomogeneities[13] and/or three-dimensional effects[16] on WASP-76b.

Beyond identifying which species are present in WASP-76b's atmosphere, we apply a high-resolution Bayesian atmospheric retrieval framework[17] to the MAROON-X data and infer the presence of a stratosphere, bounded abundance constraints for 13 elements and molecules, as well as upper limits on several other species (see Extended Data Fig. 5 and Extended Data Table 1). We find that WASP-76b reaches temperatures significantly hotter than its equilibrium temperature of $T_{eq}$ = 2,228 K assuming zero albedo[7]. The hot stratosphere is also consistent with the presence of ionized species such as $Ca^+$ and $Ba^+$ at high altitude (Extended Data Fig. 6).

Comparing the inferred elemental abundances on WASP-76b to host star[18] and proto-solar[5] values, we find that the abundances of Mn, Cr, Mg, Ni, V, Ba, and Ca all follow a remarkably similar trend, especially when taken relative to Fe (Fig. 2a). This agreement between chemically unfractionated materials and planetary relative abundances spans across several orders of magnitude and sharply contrasts from compositions of highly differentiated bodies such as Earth's crust[19], for example. This indicates that WASP-76b's present-day atmosphere, to first order, has a similar refractory composition to the parent protoplanetary disk from which it was formed, similarly to what was found for a subset of these elements on another ultra-hot Jupiter WASP-121b[3,4]. We measure the abundance of neutral alkali metals Li, Na, and K in the photosphere of WASP-76b to be significantly sub-solar, which is naturally explained by their relatively low ionization potentials causing these elements to be heavily ionized at the probed temperatures and pressures. With their respective ions ($Li^+$, $Na^+$, $K^+$) lacking spectral features due to not having any valence electrons, they cannot be remotely probed with MAROON-X. This naturally leads to their abundances being under-predicted from only considering their neutral forms and therefore not being representative of the bulk atmosphere.

The ultra-refractory elements Ti, Sc, and Al, on the other hand, are severely depleted relative to proto-solar. Unlike alkali metals, these are not expected to be significantly ionized. Instead, with their higher condensation temperatures[20] ($T_{cond}$), we conclude that a cold-trap on WASP-76b must cause these highly refractory elements to be removed from the gas phase of the upper atmosphere[21]. Indeed the measured abundance ratios relative to solar show a steep transition, with elements having condensation temperatures below ~1,550 K being roughly in line with solar and elements with $T_{cond} \geq$ 1,550 K being significantly depleted (Fig. 2b). Such a cold-trap mechanism has also been proposed on the similar ultra-hot Jupiter WASP-121b ($T_{eq}$ ~ 2,350 K) to explain the lack of Ti and TiO[3,4]. Measuring a wide range of refractory abundance ratios in other giant exoplanets progressively hotter and colder than WASP-76b will be necessary to better understand the condensation sequence of mineral species in exoplanet atmospheres and indirectly probe cloud compositions[22,23]. For example, if nucleation is efficient in hot Jupiter atmospheres, we may expect their transmission spectra to show sharp transitions as mineral species are depleted, one by one as a function of their condensation temperature[24]. Similarly, hotter planets would progressively "unlock" elements, as in the case of the similar but even warmer ultra-hot Jupiter WASP-189b ($T_{eq}$ ~ 2,650 K) which shows absorption from Ti and Sc species[13]. On the other hand, the condensation sequence of different mineral species is likely less straightforward, given that heterogeneous nucleation is strongly dependent on the availability of cloud seed particles, their nucleation rates[25], and their gravitational settling timescales[26]. Identifying and quantifying similar abundance transitions where species become depleted in relation to their condensation temperatures will also serve as an indirect probe of the nightside temperature profiles of hot Jupiters[8].

While most elements on WASP-76b are either consistent with proto-solar and stellar abundances or significantly depleted due to ionization/condensation, Cr, Ni, and V do show differences at the ~2–3σ level, even when utilizing different model parameterizations (Extended Data Fig. 7). If representative of the bulk envelope, abundance ratios that deviate from proto-solar/stellar can shed light into the composition of materials accreted during WASP-76b's formation and evolution. For example, one possibility is that WASP-76b accreted a significant proportion of differentiated, non-solar-like material throughout its history. To quantify this hypothetical scenario we use a toy model that calculates the final abundances in WASP-76b's atmosphere post-accreting a body with a given composition and mass (see Methods). We find that measured refractory elemental ratios can be reasonably well matched if, for example, WASP-76b accreted Mercury-like material with a total mass half that of Earth (see Extended Data Fig. 8). If caused by a single accretion occurrence, such an event during WASP-76b's evolution could resemble the giant collision that Jupiter has been hypothesized to have undergone to explain its diluted core[27]. However, we notably are unable to perfectly match all measured abundance ratios on WASP-76b with the scenarios explored by our model (see Methods), highlighting the importance of precisely measuring a wide range of elements to help constrain the composition of any accreted material. We also cannot rule out that the inferred non-solar elemental abundance ratios are instead the result of other physical or chemical atmospheric processes. For instance, if Ni spectral lines probe deeper, colder atmospheric layers on average, it may be that all other species are slightly more ionized, making Ni appear more abundant in comparison. Meanwhile, Cr and V could plausibly also appear underabundant due to being partially condensed[28] or partly bound in other compounds (e.g., CrO, $VO_2$, see Extended Data Fig. 6).

Finally, the MAROON-X data also show a "kinked" signal in the phase-resolved absorption of both iron and other species in WASP-76b's atmosphere, with the absorption trails being progressively more blueshifted over the first half of transit before holding steady (Fig. 3). Previous observations of WASP-76b showing such a kink in iron absorption were interpreted as iron condensing out of the atmosphere from the day to the nightside as the cooler morning terminator rotated into view[7]. Given that we detect multiple species with differing condensation temperatures that all behave similarly (see also Extended Data Fig. 9) we conclude that, instead of rainout, a global process affecting most species systematically must be responsible. We thus argue that a substantial temperature asymmetry[16] and/or unevenly distributed high-altitude optically thick clouds[29] between the east and west terminators of the planet are more likely explanations. In these scenarios, most of the signal originates from WASP-76b's east terminator, with contributions from the western limb damped due to a smaller scale height and/or clouds[30]. While condensation of iron and other species may indeed be occurring from the day to the nightside, it is unlikely to be the sole cause of the observed asymmetric absorption trails of species in WASP-76b's atmosphere.

1. Lothringer, J. D. *et al.* A New Window into Planet Formation and Migration: Refractory-to-Volatile Elemental Ratios in Ultra-hot Jupiters. *The Astrophysical Journal* **914**, 12 (2021).

2. Atreya, S. K., Mahaffy, P. R., Niemann, H. B., Wong, M. H. & Owen, T. C. Composition and origin of the atmosphere of Jupiter—an update, and implications for the extrasolar giant planets. *Planetary and Space Science* **51**, 105–112 (2003).

3. Gibson, N. P., Nugroho, S. K., Lothringer, J., Maguire, C. & Sing, D. K. Relative abundance constraints from high-resolution optical transmission spectroscopy of WASP-121b, and a fast model-filtering technique for accelerating retrievals. *Monthly Notices of the Royal Astronomical Society* (2022).

4. Maguire, C. *et al.* High-resolution atmospheric retrievals of WASP-121b transmission spectroscopy with ESPRESSO: Consistent relative abundance constraints across multiple epochs and instruments. *Monthly Notices of the Royal Astronomical Society* **519**, 1030–1048 (2023).

5. Asplund, M., Amarsi, A. M. & Grevesse, N. The chemical make-up of the Sun: A 2020 vision. *Astronomy & Astrophysics* **653**, A141 (2021).

6. Fortney, J. J., Lodders, K., Marley, M. S. & Freedman, R. S. A Unified Theory for the Atmospheres of the Hot and Very Hot Jupiters: Two Classes of Irradiated Atmospheres. *The Astrophysical Journal* **678**, 1419 (2008).

7. Ehrenreich, D. *et al.* Nightside condensation of iron in an ultrahot giant exoplanet. *Nature* 1–7 (2020).

8. Lothringer, J. D. *et al.* UV absorption by silicate cloud precursors in ultra-hot Jupiter WASP-178b. *Nature* **604**, 49–52 (2022).

9. West, R. G. *et al.* Three irradiated and bloated hot Jupiters: - WASP-76b, WASP-82b, and WASP-90b. *Astronomy & Astrophysics* **585**, A126 (2016).

10. Seifahrt, A., Stürmer, J., Bean, J. L. & Schwab, C. MAROON-X: A radial velocity spectrograph for the Gemini Observatory. in *Ground-based and Airborne Instrumentation for Astronomy VII* vol. 10702 107026D (2018).

11. Snellen, I. A. G., Kok, R. J. de, Mooij, E. J. W. de & Albrecht, S. The orbital motion, absolute mass and high-altitude winds of exoplanet HD 209458b. *Nature* **465**, 1049–1051 (2010).

12. Lothringer, J. D., Barman, T. & Koskinen, T. Extremely Irradiated Hot Jupiters: Non-oxide Inversions, H- Opacity, and Thermal Dissociation of Molecules. *The Astrophysical Journal* **866**, 27 (2018).

13. Prinoth, B. *et al.* Titanium oxide and chemical inhomogeneity in the atmosphere of the exoplanet WASP-189 b. *Nature Astronomy* 1–9 (2022).

14. Spiegel, D. S., Silverio, K. & Burrows, A. CAN TiO EXPLAIN THERMAL INVERSIONS IN THE UPPER ATMOSPHERES OF IRRADIATED GIANT PLANETS? *The Astrophysical Journal* **699**, 1487–1500 (2009).

15. Mahaffy, P. R. *et al.* Noble gas abundance and isotope ratios in the atmosphere of Jupiter from the Galileo Probe Mass Spectrometer. *Journal of Geophysical Research: Planets* **105**, 15061–15071 (2000).

16. Wardenier, J. P., Parmentier, V., Lee, E. K. H., Line, M. R. & Gharib-Nezhad, E. Decomposing the iron cross-correlation signal of the ultra-hot Jupiter WASP-76b in transmission using 3D Monte Carlo radiative transfer. *Monthly Notices of the Royal Astronomical Society* **506**, 1258–1283 (2021).

17. Pelletier, S. *et al.* Where Is the Water? Jupiter-like C/H Ratio but Strong H2O Depletion Found on tau Boötis b Using SPIRou. *The Astronomical Journal* **162**, 73 (2021).

18. Tabernero, H. M. *et al.* ESPRESSO high-resolution transmission spectroscopy of WASP-76 b. *Astronomy & Astrophysics* **646**, A158 (2021).

19. Hans Wedepohl, K. The composition of the continental crust. *Geochimica et Cosmochimica Acta* **59**, 1217–1232 (1995).
20. Lodders, K. Solar System Abundances and Condensation Temperatures of the Elements. *The Astrophysical Journal* **591**, 1220 (2003).
21. Roman, M. T. *et al.* Clouds in Three-dimensional Models of Hot Jupiters over a Wide Range of Temperatures. I. Thermal Structures and Broadband Phase-curve Predictions. *The Astrophysical Journal* **908**, 101 (2021).
22. Lothringer, J. D., Fu, G., Sing, D. K. & Barman, T. S. UV Exoplanet Transmission Spectral Features as Probes of Metals and Rainout. *The Astrophysical Journal Letters* **898**, L14 (2020).
23. Gao, P., Wakeford, H. R., Moran, S. E. & Parmentier, V. Aerosols in Exoplanet Atmospheres. *Journal of Geophysical Research: Planets* **126**, e2020JE006655 (2021).
24. Grossman, L. Condensation in the primitive solar nebula. *Geochimica et Cosmochimica Acta* **36**, 597–619 (1972).
25. Gao, P. *et al.* Aerosol composition of hot giant exoplanets dominated by silicates and hydrocarbon hazes. *Nature Astronomy* 1–6 (2020).
26. Powell, D. *et al.* Transit Signatures of Inhomogeneous Clouds on Hot Jupiters: Insights from Microphysical Cloud Modeling. *The Astrophysical Journal* **887**, 170 (2019).
27. Liu, S.-F. *et al.* The formation of Jupiter's diluted core by a giant impact. *Nature* **572**, 355–357 (2019).
28. Morley, C. V. *et al.* NEGLECTED CLOUDS IN T AND Y DWARF ATMOSPHERES. *The Astrophysical Journal* **756**, 172 (2012).
29. Savel, A. B. *et al.* No Umbrella Needed: Confronting the Hypothesis of Iron Rain on WASP-76b with Post-processed General Circulation Models. *The Astrophysical Journal* **926**, 85 (2022).
30. Kesseli, A. Y., Snellen, I. A. G., Casasayas-Barris, N., Mollière, P. & Sánchez-López, A. An Atomic Spectral Survey of WASP-76b: Resolving Chemical Gradients and Asymmetries. *The Astronomical Journal* **163**, 107 (2022).

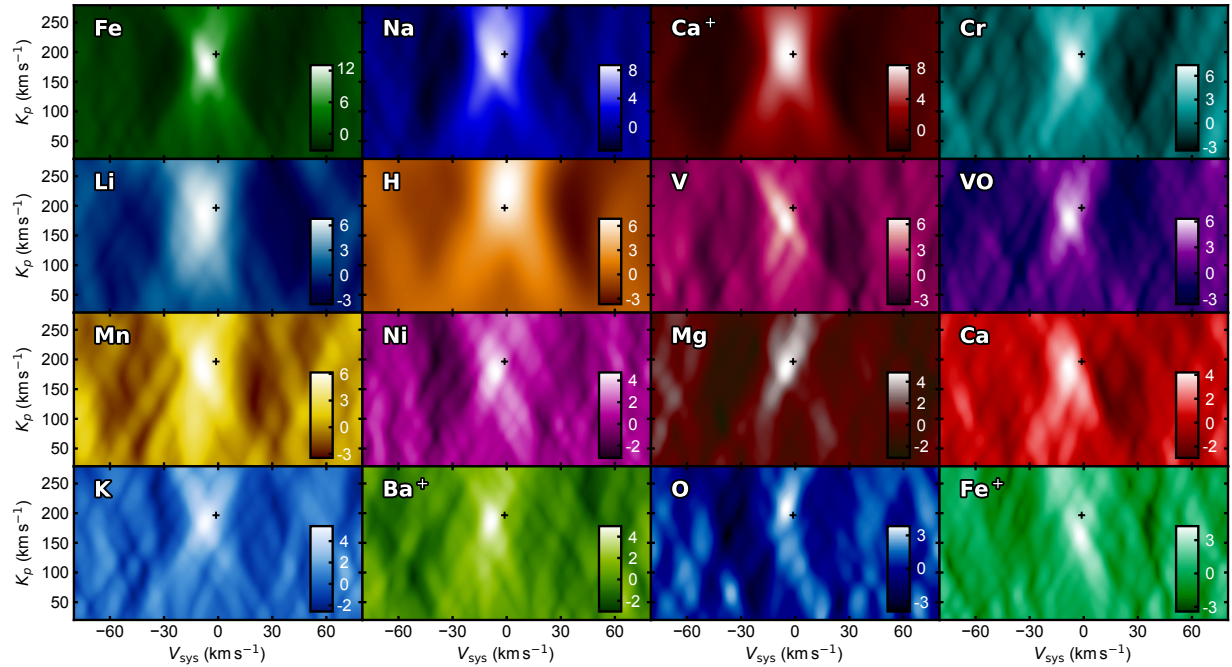

**Fig. 1 | Cross-correlation results for species detected in WASP-76b's atmosphere.** In each panel, the signal-to-noise velocity map of an element is shown, with the black cross indicating the expected location of the signal assuming a symmetric planet with a static atmosphere. Deviations from the black cross in the $K_p$–$V_{sys}$ space can be indicative of chemical asymmetries and dynamics on WASP-76b. Clear signals, many of which slightly offset, can be seen as bright white blobs near the expected position for all 16 species shown.

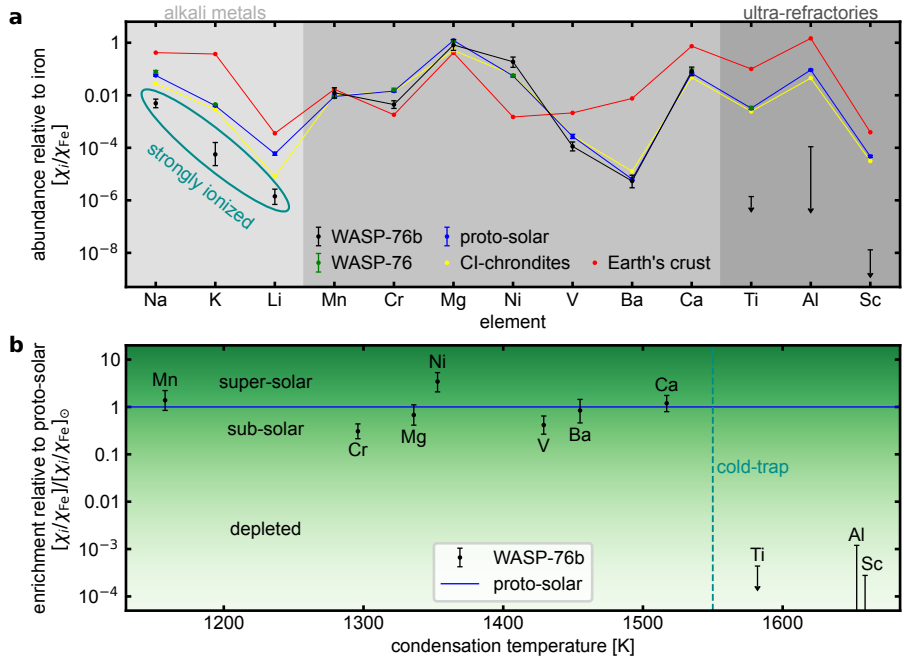

**Fig. 2 | Retrieved elemental composition of WASP-76b's atmosphere relative to iron. a**, Elemental abundance ratios relative to iron on WASP-76b compared to proto-solar[5], stellar[18], and CI-chrondrites[20] compositions. Other than alkali metals and ultra-refractories, elements in WASP-76b's atmosphere follow a strikingly similar trend as these primitive, unprocessed materials. For comparison, we also show Earth's crustal composition[19], which is highly processed and unsurprisingly poorly represents the data. Alkali metals are measured to be underabundant due to being strongly ionized at the low pressures and high temperatures probed and thus likely do not represent the true atmospheric abundance. **b**, Measured refractory abundance ratios in WASP-76b's atmosphere relative to proto-solar. WASP-76b's atmospheric enrichment is near-proto-solar (blue line) for elements with condensation temperatures up to ~1,550 K, before sharply transitioning and showing orders of magnitude depletion levels. With their higher condensation temperatures, ultra-refractory elements (Ti, Sc, Al) likely appear depleted due to being cold-trapped on the planet's colder nightside. From the near proto-solar abundance of V/Ba/Ca and severe depletion of Ti/Al/Sc, we can constrain the cold-trap temperature to be between ~1,520 and 1,580 K. All errorbars represent 1σ uncertainties.

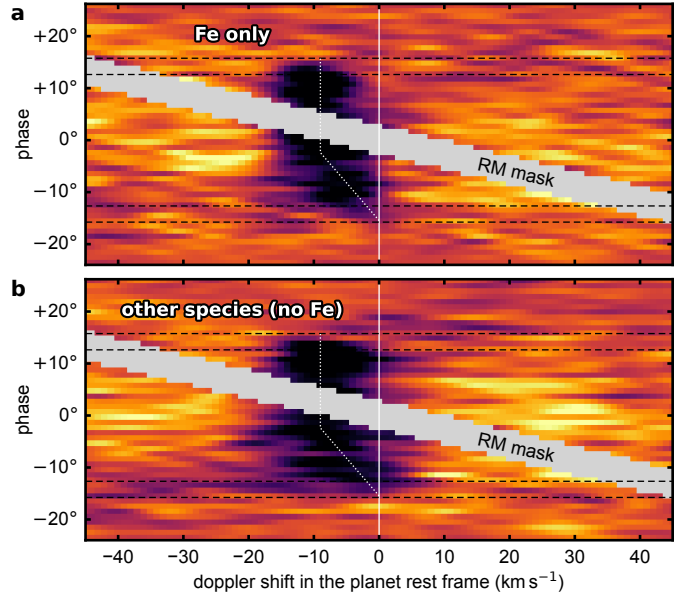

**Fig. 3 | Rest frame absorption signals on WASP-76b. a**, Cross-correlation signature of iron absorption showing a leftward "kink" (white dotted line) occurring during transit. The dark trail represents the planet's absorption signature, which is only present during the transit event. The black dashed lines represent ingress and egress contact points while the gray shaded area is a region in radial velocity space not considered in the analysis due to overlapping with the stellar Rossiter-Mclaughlin effect ("RM mask" region). A uniform and static atmosphere on WASP-76b would produce an absorption trail centered at zero. Instead, the planetary signature is kinked, starting at a Doppler shift close to zero and becoming progressively more blueshifted during the first half of the transit before holding steady. **b**, Same but for other species included in the retrieval excluding iron (Na, K, Li, Mn, Cr, Mg, Ni, V, VO, Ba, Ca, and $Ca^+$), also showing a similar asymmetric absorption trace (see Extended Data Fig. 9 for the individual contribution of each species). The asymmetry of the signal is not unique to iron, indicating that a global process affecting many elements similarly such as an uneven temperature or distribution of clouds on the eastern and western limbs of WASP-76b is causing this behavior.

# Methods

**MAROON-X observations of WASP-76b**

WASP-76b is a tidally-locked ultra-hot Jupiter orbiting the F7-type star WASP-76 ($V = 9.5$, $T_{eq} = 6329 \pm 65$ K). With a short orbital period of only 1.81 days, WASP-76b has a mass slightly less than that of Jupiter ($0.894^{+0.014}_{-0.013}$ $M_{Jup}$) but a radius that is much larger ($1.854^{+0.077}_{-0.076}$ $R_{Jup}$)[7]. We used the MAROON-X spectrograph to obtain three continuous time series observations of WASP-76b as it transited its host star. The observations were part of Gemini-North programs GN-2020B-Q-122 (principal investigator: Pelletier) and GN-2021B-Q-138 (principal investigator: Debras). MAROON-X covers the full 491 – 678 nm wavelength range across 34 spectral orders in its "blue" arm, and the full 647 – 921 nm wavelength range across 28 spectral orders in its "red" arm. As the detectors in the blue and red arms have different readout times, we integrated on the blue arm 40 s longer per exposure to maintain a constant cadence for the first two transits. The third transit maintained a fixed exposure time of 480 s for both channels. With the exception of the first transit, which misses the first 2 – 3 minutes of ingress, all three nights cover the entire transit, including start and end baseline, all under photometric sky conditions. For the analysis, we use the most up-to-date full orbital solution[7], updated with the latest ephemeris from TESS[31]. WASP-76 is part of a binary system[32], with a companion near the limit of the 0.77" field of view of MAROON-X. However, as the companion is effectively stationary in velocity space over the course of the observations, any starlight contamination is removed along with the primary star during the data detrending.

**Data reduction and detrending**

We used the standard MAROON-X pipeline[33] to extract detector images into one-dimensional wavelength-calibrated spectra, order by order for each exposure. The pipeline outputs a $N_{exp} \times N_{order} \times N_{pixel}$ high-resolution spectral time series for each channel of each transit sequence, with $N_{exp}$, $N_{order}$, and $N_{pixel}$ being the number of exposures, number of spectral orders, and number of pixels for each detector. The data contains the stellar spectrum, absorption from the transiting planet, Earth's tellurics, and noise. Our objective is to remove the stellar and telluric contributions that dominate the observed spectra, without removing the relatively much fainter planetary signal. This is possible to achieve because WASP-76b's absorption features experience a change in radial velocity of ~100 km s$^{-1}$ over the duration of a full transit, while telluric and stellar spectral lines remain relatively stationary in wavelength. We apply a PCA-based reduction algorithm to remove all unwanted wavelength-stationary contributions and uncover the atmospheric signature of WASP-76b. The data detrending procedure broadly follows ref.[17], with some minor alterations. Here we provide a summary of the steps applied to each order of the transit time series independently.

1. Outliers in the spectroscopic time series deviating by more than 5σ from the median of their spectral channel are flagged and corrected via interpolation of neighboring wavelengths.
2. All observed spectra are aligned in velocity space to remove both the Earth's barycentric motion and WASP-76's reflex motion. This procedure ensures that stellar lines are well wavelength-aligned and can be properly removed.
3. Blaze and throughput variations are removed by bringing each spectrum in the time series to the same continuum level. This consists of first dividing each pixel by its median in time, smoothing the residuals with a low-pass filter, and then dividing the original spectra by this filter[3,34,35]. The smoothing is done by passing a median box filter with a width of 501 pixels that is subsequently convolved by a Gaussian with a standard deviation of 100 pixels. The large box kernel width and Gaussian standard deviation used relative to the 1.1 km s$^{-1}$ average velocity dispersion per pixel of MAROON-X are necessary to avoid altering the exoplanet's signal[3].
4. Each spectrum is divided by a second-order polynomial fit of the median of all out-of-transit exposures.
5. A PCA reconstruction of the first 10 principal components summed together is divided out of each spectrum. The choice to remove 10 principal components is relatively arbitrary as removing anywhere between ~3–20 components does not significantly change our results. At least a few must be removed to properly eliminate stellar and telluric contributions, but too many would remove the planet's atmospheric signal.
6. Spectral channels that have a standard deviation greater than four times that of the median of their spectral order are masked and removed from the analysis.
7. An uncertainty of the form $\sigma_n = (aF_n + b)^{0.5}$ is estimated for every flux value. Here $F_n$ is the measured flux for each data point $n$ while $a$ and $b$ are fitted via a likelihood minimization of the data residuals after a PCA reduction. This is the same procedure used in refs.[3,35], to which the reader is referred to for more details.

An overview of these steps applied on an example spectral order is shown in Extended Data Fig. 1. While not all spectral orders are equal in terms of telluric and stellar line content, and may benefit from different levels of detrending, we refrain from optimizing the number of principal components removed in each order to not introduce additional degrees of freedom that could potentially bias the analysis[36,37]. The 760–770 nm wavelength range dominated by deep telluric absorption is masked out and not considered in the analysis. The telluric- and stellar-free cleaned data residuals and their associated uncertainties are then used to uncover WASP-76b's atmospheric trace hidden in the noise via cross-correlation with atmosphere models.

**Atmospheric modeling**
To properly study WASP-76b's atmosphere, it is crucial to have a representative model. For our analysis, we generate synthetic high-resolution transmission spectra using the SCARLET framework[17,38–42] with cross sections computed using HELIOS-K1[43,44]. Opacities used for this work include AlO[45], CrH[46,47], K[48], Na[49], TiO[50], VO[51], and all other atoms and ions available in either the VALD[52], Kurucz[53], or NIST[54] databases. Our main results use the VALD database, but we use Kurucz and NIST to verify the retrieved refractory abundance ratios (Extended Data Fig. 7a). The absorption cross sections for a subset of these species that are of interest for this work are shown in Extended Data Fig. 5. While the atomic cross sections do not include pressure broadening (other than Na and K), this is not likely to have a significant impact due to low (sub-millibar) pressures at which metallic lines are probed in this data set. SCARLET models also include collision-induced absorption from $H_2$–$H_2$ and $H_2$–He interactions[55], as well as Rayleigh scattering[38]. An optically thick gray cloud deck at a freely parameterized cloud-top pressure $P_c$ is also included. Although H$^-$ is likely the dominant continuum opacity source on WASP-76b, because H$^-$ opacity[56,57] is nearly flat over the MAROON-X bandpass (Extended Data Fig. 2) and therefore affects modeled spectra similarly to a cloud deck, we opt to fit for $P_c$ rather than H$^-$ due to its more easily interpretable effect on the resulting transmission spectrum. We use $H_2$, He, and H as filler gases such that the volume mixing ratio of all species included in the model always add to one in every atmospheric layer. The relative amounts of $H_2$, He, and H added depend on the temperature-pressure profile and is determined using FastChem[58], with $H_2$ typically being most abundant deeper in the atmosphere and H dominating at higher altitudes (see Extended Data Fig. 6b). Accounting for the transition between atomic and molecular hydrogen is important as it traces a significant change in mean molecular weight in the atmosphere. Models are generated at a resolution of $R = 250{,}000$ over the full MAROON-X wavelength range and later broadened to the instrumental resolution of $R = 85{,}000$.

---

[1] https://chaldene.unibe.ch/

**Cross-correlation analysis**
We perform a classical cross-correlation analysis to search for individual species in WASP-76b's atmosphere. Given cleaned data residuals $d_n$ with uncertainties $\sigma_n$ (see Extended Data Fig. 1) and a model transmission spectrum $m_n$, the cross-correlation function is defined as

$$\mathrm{CCF}(v) = \sum_n \frac{d_n m_n(v)}{\sigma_n^2},$$

and is calculated at a given radial velocity shift ($v$) for each spectrum of the time series by summing over every wavelength bin (n). Computed for a wide range of velocities and summing all spectral orders together, this produces a CCF time series where the absorption from the planet's atmosphere should follow a distinct trail along the expected orbital path[11]. One can then integrate in time by phase-folding the data, summing contributions from all exposures for different combinations of $K_p$ and $V_{\mathrm{sys}}$, effectively generating a velocity–velocity map that should produce a peak near the known position of the planet if the modeled atmosphere matches the true signal (see Fig. 1). Deviations from the expected $K_p$ and $V_{\mathrm{sys}}$ may then be indicative of dynamical or asymmetric chemical processes at play in WASP-76b's atmosphere[13]. Signal-to-noise values for a given map are computed by dividing summed CCF values by the standard deviation of all values away from the expected position of WASP-76b's signal. As in previous work for this target[30], values falling within $\pm 10\,\mathrm{km\,s^{-1}}$ of $V_{\mathrm{sys}}$ are masked out (see Fig. 3) and not included in the CCF or retrieval analyses to avoid contamination from the Rossiter-McLaughlin effect[59,60].

**Bayesian inference framework**
While a cross-correlation analysis is the method of choice for investigating whether a given element or molecule is present in the atmosphere being probed, it provides only limited quantitative information on their abundances. We detect iron on WASP-76b for example, but how much iron is there? To further characterize WASP-76b's atmosphere we perform a Bayesian high-resolution retrieval analysis using the likelihood prescription of refs.[3,35] given by

$$\ln(L) = -\frac{N}{2}\ln\left(\frac{1}{N}\sum_n \frac{(d_n - \alpha m_n)^2}{\sigma_n^2}\right),$$

where $N$ is the total number of data points, and $\alpha$ is a model scaling parameter. We do not include a noise scaling term $\beta$, which has a negligible impact on large, well-behaved data sets[35].

To accurately infer atmospheric parameters, we mimic the effects that both WASP-76b's rotation and the reduction steps applied to the data have on the underlying signal to the model. For this, the model $m_n$ is convolved with a rotational broadening kernel assuming tidally-locked rotation, projected in time for a given combination of $K_p$ and $V_{sys}$, and multiplied by a stellar quadratic limb darkening model. We follow the approach outlined in ref.[61] to compute the limb darkening model (see their equations 16 – 22), but assume the planet to be uniform. We use an impact parameter $b$ = 0.027 and quadratic limb darkening coefficients $\mu_1$ = 0.393 and $\mu_2$ = 0.219 (ref.[7]). The model is then injected in the removed PCA reconstruction of the data (step 5 of the data reduction), and the PCA algorithm is run once again on this mock data set including the injected model[17]. The output is a modified $m_n$ better representative of what we would expect the true signal to be after the application of the detrending algorithm. This modified $m_n$ is used as an input to equation 1 and ensures that accurate parameters can be inferred[62,63].

For the retrieval, we fit the log volume mixing ratio ($\log_{10}\chi$, prior between −14 and 0) of all 20 species simultaneously, making no a priori assumptions on the chemistry and instead setting abundances to be uniform in pressure. We also simultaneously fit for the temperature structure ($T_0$–$T_9$, prior between 0 and 7000 K), log continuum pressure ($\log_{10}P_c$, prior between −8 and 2 bars), log scale factor ($\log_{10}\alpha$, prior between −2 and 2), and orbital and systemic velocities $K_p$ (prior between 166.5 and 226.5 km s$^{-1}$) and $V_{sys}$ (prior between −21.1 and 18.9 km s$^{-1}$). The temperature-pressure profile is freely fit as parametrized in ref.[17] (see their equation 11), using ten temperature points uniformly distributed in log pressure between $10^{-10}$ ($T_0$) and 1 ($T_9$) bars, and a smoothing prior $\sigma_s$ = 160 K dex$^{-2}$. All parameters are fit simultaneously to ensure that all degeneracies are marginalized over in the parameter inference[62]. The Bayesian inference and parameter exploration is done using the emcee[64] Markov Chain Monte Carlo (MCMC) python package. We run the MCMC for 30,000 steps with 200 walkers and use the last 10,000 steps (2,000,000 samples) to compute posterior distributions. The retrievals are computationally expensive, needing approximately three months to run when using 18 CPU cores in parallel. We test convergence by employing the Gelmaan–Rubin diagnostic for each chain. We also supplement our retrieval with the white light curve transit depth from existing HST/STIS data[65] to anchor models at the correct planetary radius.

To estimate total elemental abundances in WASP-76b's atmosphere, vanadium is calculated from the sum of V and VO, titanium from Ti and TiO, calcium from Ca and Ca$^+$, chromium from Cr and CrH, aluminium from Al and AlO, barium from Ba$^+$, and other species from their neutral forms (e.g., magnesium from Mg). In most cases, the species included in the retrieval make up the dominant portion of the expected chemical inventory at the pressures probed (Extended Data Fig. 6). We note that a limitation of our atmospheric model is that it does not account for abundance variations in pressure due to ionization, which can potentially bias inferred volume mixing ratios, as is evidently the case for the alkali metals. For example, iron may appear less abundant than in reality due to partial ionization, especially given the tentative Fe$^+$ signal observed in the data – although this signal likely originates from higher altitudes than probed by neutral Fe. Similarly, retrieved abundances might be underestimated due to partial condensation, or from not considering additional species. For example, V and Ti may be partly bound in VO$_2$ and TiO$_2$ molecules that we cannot measure due to the lack of suitable line lists[66].

We note that the retrieved velocity parameters are driven by the strongest absorber (i.e., Fe), and that other species can have small relative offsets in $K_p$ and $V_{sys}$. However, we refrain from fitting $K_p$ and $V_{sys}$ freely for each species in order to obtain meaningful abundance ratios. The extreme examples, H, O, and Fe$^+$, are excluded from the retrieval as their peaks are significantly offset from the other detected species (Fig. 1), potentially due to strong vertical winds or outflows[30,59].

Similarly, some absorption features probe up to non-hydrostatic layers of the atmosphere[4,18] (e.g., $Ca^+$) and may not be well represented by our model, which could bias the abundances we retrieve. We also note that our retrieval is based on a one-dimensional atmospheric model and only captures average properties of the inherently three-dimensional planet, which could result in further biases of inferred parameters[67]. However, ref.[61] show that the retrieved iron abundance on WASP-76b from one-dimensional retrievals is roughly consistent with results from two-dimensional retrievals. Nevertheless, we chose not to over-interpret retrieved absolute abundances and rather focus on relative abundance measurements which can be constrained much more precisely and are less sensitive to systematic biases[3].

We recover orbital velocity parameters $K_p$ = 180.7 ± 0.6 km s$^{-1}$ and $V_{sys}$ = -7.0 ± 0.1 km s$^{-1}$. The discrepancy between these and the known literature values is due the planetary rotation, day-to-night atmospheric winds, and an artefact of fitting a kinked absorption trail with a traditional planetary orbit and should be interpreted as an indicator of an asymmetry in WASP-76b's eastern and western limbs[7]. Similarly, the retrieved continuum pressure level (in bars) of $\log_{10} P_c = -2.21^{+0.27}_{-0.28}$ and temperature profile are an average of both hemispheres, when in reality this value is likely different on each terminator[61]. We also recover a scaling factor of $\alpha = 0.52^{+0.08}_{-0.09}$, which may be decreasing the amplitude of the models to compensate for an evening terminator signal that is damped due to a colder temperature and/or clouds[16,29,30]. The retrieved temperature profile is consistent with previous findings of WASP-76b's upper atmosphere being hotter than expected from model predictions[59,61,68,69]. We estimate WASP-76b's total refractory metallicity (i.e., the sum of all retrieved absolute abundances) to be $\log_{10}(\Sigma \chi_i) = -3.45^{+0.30}_{-0.32}$, slightly enriched relative to both solar[5] (+0.62 dex) and stellar[18] (+0.28 dex) compositions (Extended Data Fig. 5b). We further test the robustness of our results by performing separate retrievals using both different line lists and temperature structure parameterizations. We find that regardless of the atomic line lists used (VALD[52], Kurucz[53], or NIST[54]) and assumed temperature-pressure profile form (free[17], Guillot[70], or isothermal), the inferred elemental abundance ratios are all relatively consistent (Extended Data Fig. 7a). However, we caution that fitted temperature profiles with less flexibility can lead to the shape of the temperature structure being over-constrained (Extended Data Fig. 7b).

**Notable non-detections**
We notably do not detect Ti, TiO, Ti$^+$, V$^+$, Sc, CrH, AlO, or Al (Extended Data Fig. 3). A previous analysis of ESPRESSO data[30] has also shown Sr$^+$ and Co detections on WASP-76b, however, MAROON-X does not cover the bluer wavelengths at which these species have most of their absorption features. We therefore cannot confirm or exclude their presence in WASP-76b's atmosphere due to lack of sensitivity. On the other hand, MAROON-X can very effectively probe Ti, TiO, Sc, and AlO. The absence of detections and derived strict upper limits for these species indicate that they are indeed depleted from WASP-76b's upper atmosphere. Given that Ti, Sc, and Al compounds all appear to be heavily underabundant, and that these elements have a higher condensation temperature than other metals that are readily detected, we concluded that these species must be condensed out of the upper atmosphere and be cold-trapped on the nightside.

**Vanadium oxide detection**
The detection of vanadium oxide (VO) on WASP-76b is the first time this molecule is unambiguously observed using high-resolution spectroscopy on an exoplanet. VO, along with TiO, have long been theorized to be drivers for thermal inversions in ultra-hot Jupiter atmospheres due to the very strong opacity of these molecules at optical wavelengths[6,71] (see also Extended Data

Fig. 2). Despite this interest, VO has proven to be notoriously difficult to robustly detect in exoplanet atmospheres. For example, VO was tentatively observed on a different ultra-hot Jupiter, WASP-121b[72–74], but additional observations later refuted this claim[75–77]. The ambiguity arises because at low resolution the broad absorption bands of VO can be hard to distinguish from other opacity sources, or from systematic noise in the data[34,75]. High-resolution spectroscopy can mitigate these difficulties, but is also more sensitive to the accuracy of the theoretical opacities used to generate the modeled spectra[78]. Due to the most up-to-date high-temperature vanadium oxide line list being known to have inaccuracies in some wavelength ranges[79,80], we confirm our results using an independent analysis method applied on a different data set. We analyzed existing WASP-76b ESPRESSO[81] transit data[7] (programme 1102.C-744) using the same methodology outlined in ref.[13] and also recover a clear VO detection (Extended Data Fig. 4b). This means that the VO signal on WASP-76b is especially strong such that we are able to detect it, even with an imperfect line list. We caution that our inferred abundance of VO is only as good as the cross sections used, which may bias our results[62]. We note that a new VO line list is currently being computed by the ExoMol group (J. Tennyson, priv. comm.), which once released may increase our capabilities at detecting VO on other, less favorable exoplanets.

**Accretion model**

We explore theoretical scenarios where WASP-76b's atmospheric composition may have been enriched by the accretion of non-solar, differentiated materials. For this, we build a chemical composition toy model which takes the following parameters as input: WASP-76b's envelope mass, its pre-merger volatile and refractory abundances, and the accreted body's core and mantle masses and their chemical composition. The model then outputs the predicted post-merger atmospheric composition, assuming the accreted body is well mixed into WASP-76b's envelope, and to the low pressures probed by these data (e.g., via strong winds both predicted by GCM models[82], and observationally inferred[59] to exist on WASP-76b). The chemistry of the accreted object's core is sampled from the range of abundances found in iron meteorites[83], which are thought to be representative of Mercury's core composition[84]. For the mantle, we explore different chemical compositions, namely those of CI chondrites[20], CB meteorites[85], and Mercury's surface[84]. Based on a thorough exploration of parameter space, we find no single model that perfectly matches all of the observational data. Models accreting at least ~0.1 – 0.5 $M_\oplus$ ($M_\oplus$ being the mass of the Earth) of purely iron-meteorite-like material (resembling Mercury's core) can fit well Cr, Mg, Ni, V, and Ba, but not Mn, or Ca. Adding material representative of Mercury's surface on top allows us to fit Ca as well without losing the other good fits, but not Mn whose abundance in the accreted materials is always too low and thus is never fitted. Keeping the iron core but accreting a solar-composition mantle instead is practically identical to the pure-core case. The assumed initial C/O ratio and metallicity of the parent body atmosphere only affect the mass needed to be accreted to match observed abundance ratios and otherwise have no impact on the final composition. The accreted mass is inversely correlated with the abundance of Ni in the accreted material, where assuming a high Ni abundance decreases the total mass necessary to obtain the same enrichment levels. We note that while our model assumes WASP-76b's enrichment to have come from a single accreted body, other scenarios where the equivalent material instead originated from upwards dredging of a dilute core[86,87] or an influx of smaller comets and meteorites are similarly plausible.

**Data availability**

The MAROON-X data used in this work is available here: https://udemontreal-my.sharepoint.com/:f:/g/personal/stefan_pelletier_umontreal_ca/EkYThK-JMKFHlclyx7RnlIABySi6V60HuZC0c_9m6LfE6Q?e=vGErBT. The ESPRESSO data used to confirm the vanadium oxide detection is publicly available on Dace (https://dace.unige.ch/dashboard/).

**Code availability**

The MAROON-X reduction pipeline[33] used by the instrument team to perform the data extraction is public software available from Gemini at (https://github.com/GeminiDRSoftware/MAROONXDR). The atmospheric modeling and retrievals utilize SCARLET[17,40], HELIOS-K[44] (https://helios-k.readthedocs) and Fastchem[58] (https://github.com/exoclime/FastChem), emcee[64] (https://emcee.readthedocs.io/en/stable/), and corner.py[88] (https://corner.readthedocs.io/en/latest/). The ESPRESSO data analysis was performed using Tayph[66] (https://github.com/Hoeijmakers/tayph). The main analysis routines written for this work and utilizing the astropy[89,90], matplotlib[91], numpy[92], scipy[93], and scikit-learn[94] python libraries are available here: https://udemontreal-my.sharepoint.com/:f:/g/personal/stefan_pelletier_umontreal_ca/EmXMwsPp2JFCnckKJNWkf7ABrEomi5EqmadxK4Hofd7ItQ?e=73GbIp.

31. Ivshina, E. S. & Winn, J. N. TESS Transit Timing of Hundreds of Hot Jupiters. *The Astrophysical Journal Supplement Series* **259**, 62 (2022).
32. Fu, G. *et al.* The Hubble PanCET Program: Transit and Eclipse Spectroscopy of the Strongly Irradiated Giant Exoplanet WASP-76b. *The Astronomical Journal* **162**, 108 (2021).
33. Seifahrt, A. *et al.* On-sky commissioning of MAROON-X: A new precision radial velocity spectrograph for Gemini North. in *Ground-based and Airborne Instrumentation for Astronomy VIII* vol. 11447 305–325 (2020).
34. Gibson, N. P. *et al.* Revisiting the potassium feature of WASP-31b at high resolution. *Monthly Notices of the Royal Astronomical Society* **482**, 606–615 (2019).
35. Gibson, N. P. *et al.* Detection of Fe i in the atmosphere of the ultra-hot Jupiter WASP-121b, and a new likelihood-based approach for Doppler-resolved spectroscopy. *Monthly Notices of the Royal Astronomical Society* **493**, 2215–2228 (2020).
36. Cabot, S. H. C., Madhusudhan, N., Hawker, G. A. & Gandhi, S. On the robustness of analysis techniques for molecular detections using high-resolution exoplanet spectroscopy. *Monthly Notices of the Royal Astronomical Society* **482**, 4422–4436 (2019).
37. Zhang, M., Chachan, Y., Kempton, E. M.-R., Knutson, H. A. & Chang, W. (Happy). PLATON II: New Capabilities and a Comprehensive Retrieval on HD 189733b Transit and Eclipse Data. *The Astrophysical Journal* **899**, 27 (2020).
38. Benneke, B. & Seager, S. ATMOSPHERIC RETRIEVAL FOR SUPER-EARTHS: UNIQUELY CONSTRAINING THE ATMOSPHERIC COMPOSITION WITH TRANSMISSION SPECTROSCOPY. *The Astrophysical Journal* **753**, 100 (2012).

39. Benneke, B. & Seager, S. HOW TO DISTINGUISH BETWEEN CLOUDY MINI-NEPTUNES AND WATER/VOLATILE-DOMINATED SUPER-EARTHS. *The Astrophysical Journal* **778**, 153 (2013).
40. Benneke, B. Strict Upper Limits on the Carbon-to-Oxygen Ratios of Eight Hot Jupiters from Self-Consistent Atmospheric Retrieval. *arXiv:1504.07655 [astro-ph]* (2015).
41. Benneke, B. *et al.* A sub-Neptune exoplanet with a low-metallicity methane-depleted atmosphere and Mie-scattering clouds. *Nature Astronomy* **3**, 813–821 (2019).
42. Benneke, B. *et al.* Water Vapor and Clouds on the Habitable-zone Sub-Neptune Exoplanet K2-18b. *The Astrophysical Journal* **887**, L14 (2019).
43. Grimm, S. L. & Heng, K. HELIOS-K: AN ULTRAFAST, OPEN-SOURCE OPACITY CALCULATOR FOR RADIATIVE TRANSFER. *The Astrophysical Journal* **808**, 182 (2015).
44. Grimm, S. L. *et al.* HELIOS-K 2.0 Opacity Calculator and Open-source Opacity Database for Exoplanetary Atmospheres. *The Astrophysical Journal Supplement Series* **253**, 30 (2021).
45. Patrascu, A. T., Yurchenko, S. N. & Tennyson, J. ExoMol molecular line lists – IX. The spectrum of AlO. *Monthly Notices of the Royal Astronomical Society* **449**, 3613–3619 (2015).
46. Burrows, A., Ram, R. S., Bernath, P., Sharp, C. M. & Milsom, J. A. New CrH Opacities for the Study of L and Brown Dwarf Atmospheres. *The Astrophysical Journal* **577**, 986 (2002).
47. Bernath, P. F. MoLLIST: Molecular Line Lists, Intensities and Spectra. *Journal of Quantitative Spectroscopy and Radiative Transfer* **240**, 106687 (2020).
48. Allard, N. F., Spiegelman, F. & Kielkopf, J. F. K–H2 line shapes for the spectra of cool brown dwarfs. *Astronomy & Astrophysics* **589**, A21 (2016).
49. Allard, N. F., Spiegelman, F., Leininger, T. & Molliere, P. New study of the line profiles of sodium perturbed by H2. *Astronomy & Astrophysics* **628**, A120 (2019).
50. McKemmish, L. K. *et al.* ExoMol molecular line lists – XXXIII. The spectrum of Titanium Oxide. *Monthly Notices of the Royal Astronomical Society* **488**, 2836–2854 (2019).
51. McKemmish, L. K., Yurchenko, S. N. & Tennyson, J. ExoMol line lists – XVIII. The high-temperature spectrum of VO. *Monthly Notices of the Royal Astronomical Society* **463**, 771–793 (2016).
52. Ryabchikova, T. *et al.* A major upgrade of the VALD database. *Physica Scripta* **90**, 054005 (2015).
53. Kurucz, R. L. Including all the lines: Data releases for spectra and opacities. *Canadian Journal of Physics* **95**, 825–827 (2017).
54. Kramida, A., Ralchenko, Y., Reader, J. & NIST, A. T. Atomic Spectra Database. *NIST* (2009).
55. Borysow, A. Collision-induced absorption coefficients of H2 pairs at temperatures from 60 K to 1000 K. *Astronomy & Astrophysics* **390**, 779–782 (2002).
56. Bell, K. L. & Berrington, K. A. Free-free absorption coefficient of the negative hydrogen ion. *Journal of Physics B: Atomic and Molecular Physics* **20**, 801–806 (1987).
57. John, T. L. Continuous absorption by the negative hydrogen ion reconsidered. *Astronomy and Astrophysics* **193**, 189–192 (1988).
58. Stock, J. W., Kitzmann, D. & Patzer, A. B. C. FastChem 2 : An improved computer program to determine the gas-phase chemical equilibrium composition for arbitrary element distributions. *Monthly Notices of the Royal Astronomical Society* **517**, 4070–4080 (2022).
59. Seidel, J. V. *et al.* Into the storm: Diving into the winds of the ultra-hot Jupiter WASP-76 b with HARPS and ESPRESSO. *Astronomy & Astrophysics* **653**, A73 (2021).

60. Casasayas-Barris, N. *et al.* The atmosphere of HD 209458b seen with ESPRESSO - No detectable planetary absorptions at high resolution. *Astronomy & Astrophysics* **647**, A26 (2021).
61. Gandhi, S. *et al.* Spatially resolving the terminator: Variation of Fe, temperature, and winds in WASP-76 b across planetary limbs and orbital phase. *Monthly Notices of the Royal Astronomical Society* **515**, 749–766 (2022).
62. Brogi, M. & Line, M. R. Retrieving Temperatures and Abundances of Exoplanet Atmospheres with High-resolution Cross-correlation Spectroscopy. *The Astronomical Journal* **157**, 114 (2019).
63. Line, M. R. *et al.* A solar C/O and sub-solar metallicity in a hot Jupiter atmosphere. *Nature* **598**, 580–584 (2021).
64. Foreman-Mackey, D., Hogg, D. W., Lang, D. & Goodman, J. Emcee: The MCMC Hammer. *Publications of the Astronomical Society of the Pacific* **125**, 306 (2013).
65. Essen, C. von *et al.* HST/STIS transmission spectrum of the ultra-hot Jupiter WASP-76 b confirms the presence of sodium in its atmosphere. *Astronomy & Astrophysics* **637**, A76 (2020).
66. Hoeijmakers, H. J. *et al.* Hot Exoplanet Atmospheres Resolved with Transit Spectroscopy (HEARTS) - IV. A spectral inventory of atoms and molecules in the high-resolution transmission spectrum of WASP-121 b. *Astronomy & Astrophysics* **641**, A123 (2020).
67. Pluriel, W. *et al.* Toward a multidimensional analysis of transmission spectroscopy - II. Day-night-induced biases in retrievals from hot to ultrahot Jupiters. *Astronomy & Astrophysics* **658**, A42 (2022).
68. Landman, R. *et al.* Detection of OH in the ultra-hot Jupiter WASP-76b. *Astronomy & Astrophysics* **656**, A119 (2021).
69. May, E. M. *et al.* Spitzer Phase-curve Observations and Circulation Models of the Inflated Ultrahot Jupiter WASP-76b. *The Astronomical Journal* **162**, 158 (2021).
70. Guillot, T. On the radiative equilibrium of irradiated planetary atmospheres. *Astronomy & Astrophysics* **520**, A27 (2010).
71. Hubeny, I., Burrows, A. & Sudarsky, D. A Possible Bifurcation in Atmospheres of Strongly Irradiated Stars and Planets. *The Astrophysical Journal* **594**, 1011 (2003).
72. Evans, T. M. *et al.* DETECTION OF $H_2O$ AND EVIDENCE FOR TiO/VO IN AN ULTRA-HOT EXOPLANET ATMOSPHERE. *The Astrophysical Journal* **822**, L4 (2016).
73. Evans, T. M. *et al.* An ultrahot gas-giant exoplanet with a stratosphere. *Nature* **548**, 58–61 (2017).
74. Evans, T. M. *et al.* An Optical Transmission Spectrum for the Ultra-hot Jupiter WASP-121b Measured with the Hubble Space Telescope. *The Astronomical Journal* **156**, 283 (2018).
75. Mikal-Evans, T. *et al.* An emission spectrum for WASP-121b measured across the 0.8–1.1um wavelength range using the Hubble Space Telescope. *Monthly Notices of the Royal Astronomical Society* **488**, 2222–2234 (2019).
76. Mikal-Evans, T. *et al.* Confirmation of water emission in the dayside spectrum of the ultrahot Jupiter WASP-121b. *Monthly Notices of the Royal Astronomical Society* **496**, 1638–1644 (2020).
77. Wilson, J. *et al.* Gemini/GMOS optical transmission spectroscopy of WASP-121b: Signs of variability in an ultra-hot Jupiter? *Monthly Notices of the Royal Astronomical Society* **503**, 4787–4801 (2021).
78. Gandhi, S. *et al.* Molecular cross-sections for high-resolution spectroscopy of super-Earths, warm Neptunes, and hot Jupiters. *Monthly Notices of the Royal Astronomical Society* **495**, 224–237 (2020).

79. Merritt, S. R. *et al.* Non-detection of TiO and VO in the atmosphere of WASP-121b using high-resolution spectroscopy. *Astronomy & Astrophysics* **636**, A117 (2020).
80. Regt, S. de, Kesseli, A. Y., Snellen, I. a. G., Merritt, S. R. & Chubb, K. L. A quantitative assessment of the VO line list: Inaccuracies hamper high-resolution VO detections in exoplanet atmospheres. *Astronomy & Astrophysics* **661**, A109 (2022).
81. Pepe, F. *et al.* ESPRESSO at VLT - On-sky performance and first results. *Astronomy & Astrophysics* **645**, A96 (2021).
82. Beltz, H. *et al.* Magnetic Drag and 3D Effects in Theoretical High-resolution Emission Spectra of Ultrahot Jupiters: The Case of WASP-76b. *The Astronomical Journal* **164**, 140 (2022).
83. Scott, E. R. D. Iron Meteorites: Composition, Age, and Origin. *Oxford Research Encyclopedia of Planetary Science* (2020).
84. Nittler, L. R., Chabot, N. L., Grove, T. L. & Peplowski, P. N. The Chemical Composition of Mercury. in *Mercury: The View after MESSENGER* (eds. Anderson, B. J., Nittler, L. R. & Solomon, S. C.) 30–51 (2018).
85. Binzel, R. P., Walker, R. M. & Cameron, A. G. W. Systematics and Evaluation of Meteorite Classification. *Meteorites and the Early Solar System II* (2006).
86. Wahl, S. M. *et al.* Comparing Jupiter interior structure models to Juno gravity measurements and the role of a dilute core. *Geophysical Research Letters* **44**, 4649–4659 (2017).
87. Debras, F. & Chabrier, G. New Models of Jupiter in the Context of Juno and Galileo. *The Astrophysical Journal* **872**, 100 (2019).
88. Foreman-Mackey, D. Corner.py: Scatterplot matrices in Python. *Journal of Open Source Software* **1**, 24 (2016).
89. Astropy Collaboration, T. *et al.* Astropy: A community Python package for astronomy. *Astronomy & Astrophysics* **558**, A33 (2013).
90. Astropy Collaboration, T. *et al.* The Astropy Project: Building an Open-science Project and Status of the v2.0 Core Package. *The Astronomical Journal* **156**, 123 (2018).
91. Hunter, J. D. Matplotlib: A 2D Graphics Environment. *Computing in Science Engineering* **9**, 90–95 (2007).
92. Harris, C. R. *et al.* Array programming with NumPy. *Nature* **585**, 357–362 (2020).
93. Virtanen, P. *et al.* SciPy 1.0: Fundamental algorithms for scientific computing in Python. *Nature Methods* **17**, 261–272 (2020).
94. Pedregosa, F. *et al.* Scikit-learn: Machine Learning in Python. *The Journal of Machine Learning Research* **12**, 2825–2830 (2011).

**Acknowledgements** This work is based on observations obtained at the Gemini North Observatory which is operated from the summit of Maunakea. The observations at the Gemini Telescope were performed with care and respect from the summit of Maunakea which is a significant cultural and historic site. This research has made use of NASA's Astrophysics Data System and the NASA Exoplanet Archive, which is operated by the California Institute of Technology, under contract with NASA within the Exoplanet Exploration Program. S.P. is supported by the Technologies for Exo-Planetary Science (TEPS) Natural Sciences and Engineering Research Council of Canada (NSERC) CREATE Trainee Program. B.B. acknowledges funding by the NSERC, and the Fonds de Recherche du Québec – Nature et Technologies (FRQNT). M.A.-D. is supported by Tamkeen under the NYU Abu Dhabi Research

Institute, United Arab Emirates grant CAP3. B.P. acknowledges partial financial support from The Fund of the Walter Gyllenberg Foundation. D.K., A.S., and J.L.B. acknowledge funding from the David and Lucile Packard Foundation, the Heising-Simons Foundation, the Gordon and Betty Moore Foundation, the Gemini Observatory, the NSF (award number 2108465), and NASA (grant numbers 80NSSC22K0117 and 80NSSC19K0293). F.D. thanks the CNRS/INSU Programme National de Planétologie (PNP) and Programme National de Physique Stellaire (PNPS) for funding support. B.K. acknowledges funding from the European Research Council under the European Union's horizon 2022 research and innovation programme (grant agreement No 865624, GPRV) O.L. acknowledges financial support from the FRQNT [270853 & 303926], the NSERC, the Trottier Institute for Research on Exoplanets (iREx), and from the University of Montreal. A.C. acknowledges funding from the French ANR under contract number ANRCE31 (SPlaSH). This work is supported by the French National Research Agency in the framework of the Investissements d'Avenir program (ANR-15-IDEX-02), through the funding of the "Origin of Life" project of the Grenoble-Alpes University.

**Author contributions** S.P., B.B., and L.P. conceived the project. S.P. wrote the original MAROON-X observing proposal, the manuscript, and carried out the analysis of the MAROON-X data with B.B. and H.J.H. providing guidance. M.A-D. did the accretion modeling portion of the analysis. B.P. independently analyzed the ESPRESSO data to confirm the VO detection. D.K., A.S., J.B., and J.S. assisted with the observational setup, carried out the observations, and performed the MAROON-X data extraction. F.D., B.K., T.H., and A.C. acquired and contributed additional data for the project. L.B. implemented the FastChem equilibrium chemistry code in the modeling framework. A.Y.K., O.L., and N.C.-B. contributed to the stellar contamination detrending algorithm. All co-authors provided comments and suggestions about the manuscript.

**Competing interests** The authors declare no competing interests.

**Additional information**

**Supplementary information** is available for this paper at

**Correspondence and requests for materials** should be addressed to Stefan Pelletier.

**Peer review information**

**Reprints and permissions information** is available at www.nature.com/reprints.

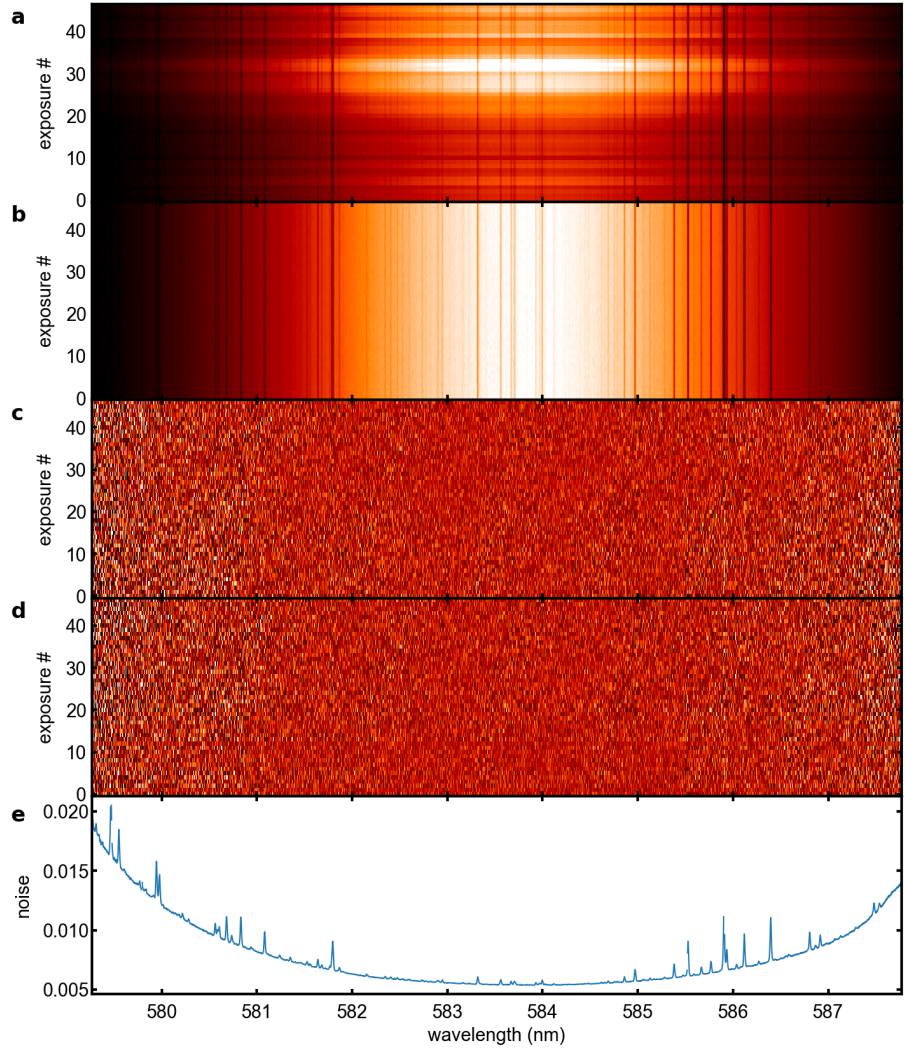

**Extended Data Figure 1 | Data detrending steps and noise model. a**, A single MAROON-X spectral order for the 5.3-hour transit time series obtained on 2020 September 12. **b**, After continuum alignment. **c**, With the median out-of-transit spectrum divided out. **d**, Post removal of stellar and telluric contaminants using PCA, with ten principal components removed in this case. **e**, The noise model used for an example exposure. The reduction steps take care of removing all telluric and stellar lines as well as the continuum while the noise model serves to downweigh regions of the spectrum that are noisier. The end product of the data reduction (panel d) contains the planetary trace buried in noise and serves as the input for the later cross-correlation and retrieval analyses.

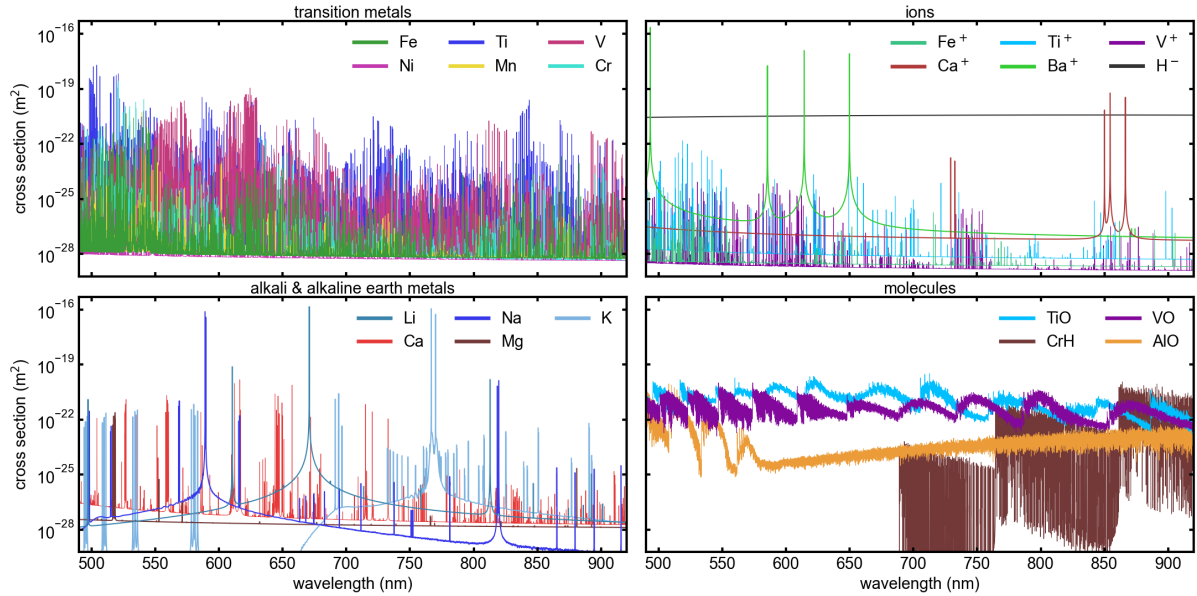

**Extended Data Figure 2 | Cross sections of prominent metals, ions, and molecules.** Values are computed for a temperature of 2500 K and include $10^{-4}$ bar pressure broadening where applicable. Atoms with only a single valence electron (Li, Na, K, $Ca^+$, and $Ba^+$) have few very strong lines while most other metals tend to be composed of "line forests" on top of a low continuum. Molecules also have forests of spectral lines, but on top of distinct absorption bands. Meanwhile, $H^-$ acts as a flat source of continuum opacity in this wavelength range, similar to a gray cloud deck. The product of a species' cross section and volume mixing ratio in WASP-76b's atmosphere determines how detectable it is.

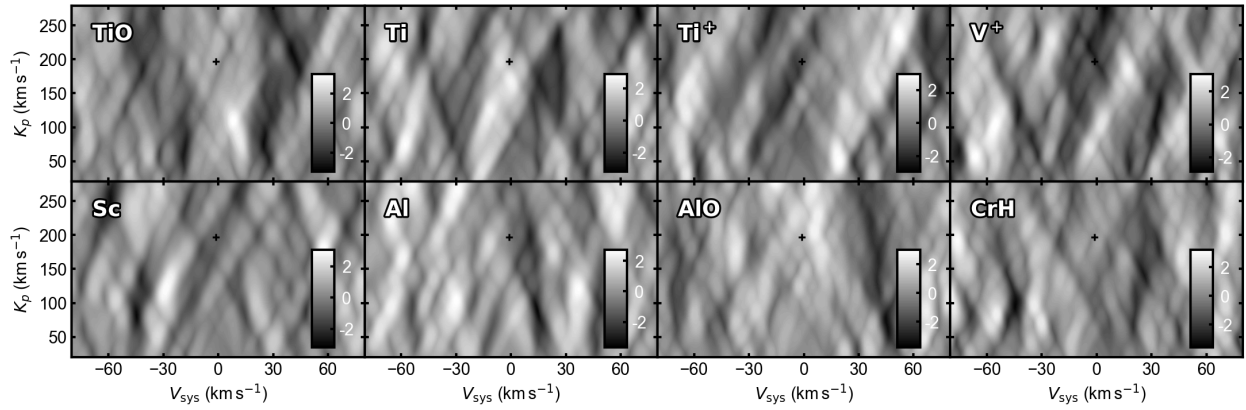

**Extended Data Figure 3 | Cross-correlation results for species not detected in WASP-76b's atmosphere.** Same as Figure 1, but for species of interest that were notably not detected. Particularly striking is the absence of Ti, TiO, Sc, and AlO which MAROON-X should be able to easily detect. Other non-detections are less surprising as they either have few strong lines in the MAROON-X bandpass (Al), or are not expected to be particularly abundant from chemical equilibrium predictions ($Ti^+$, $V^+$, CrH).

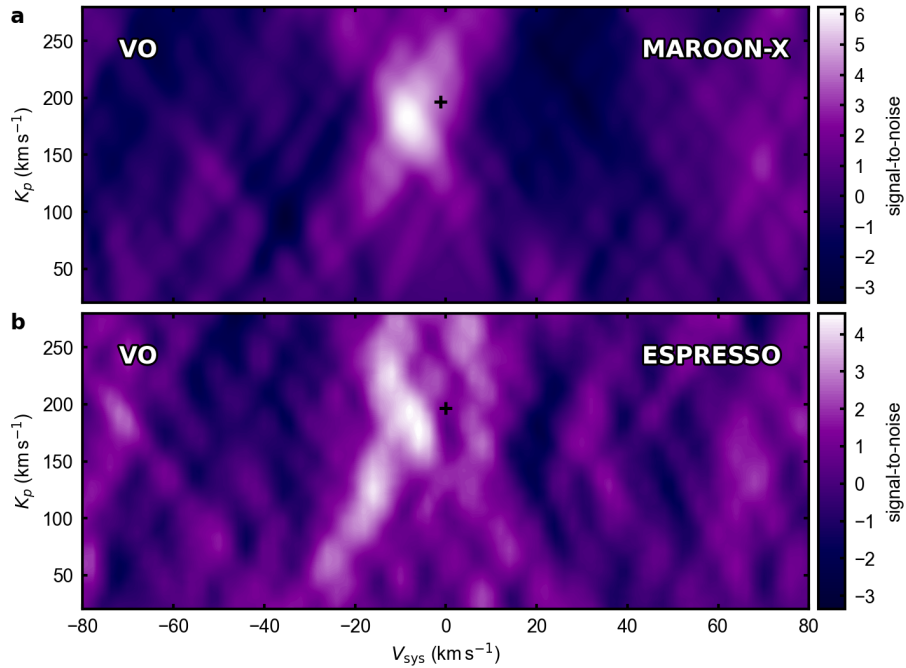

**Extended Data Figure 4 | Independent confirmation of the VO detection on WASP-76b. a**, The vanadium oxide detection with MAROON-X (490 – 920 nm, $R$ = 85,000, 3 transits). **b**, The VO detection with ESPRESSO (380 – 790 nm, $R$ = 140,000, 2 transits)[7]. The ESPRESSO analysis was done with an independent framework[13], using an independent model template, on a separate data set, and also shows a clear VO signal slightly offset to the expected location (black cross) that is consistent with the MAROON-X data. Differences in the overall detection strength and shape are due to MAROON-X having a lower resolution, but a redder wavelength coverage, and three instead of two transits. The presence of a VO signal in both the MAROON-X and ESPRESSO transit data sets provides a high degree of confidence as to the robustness of the signal.

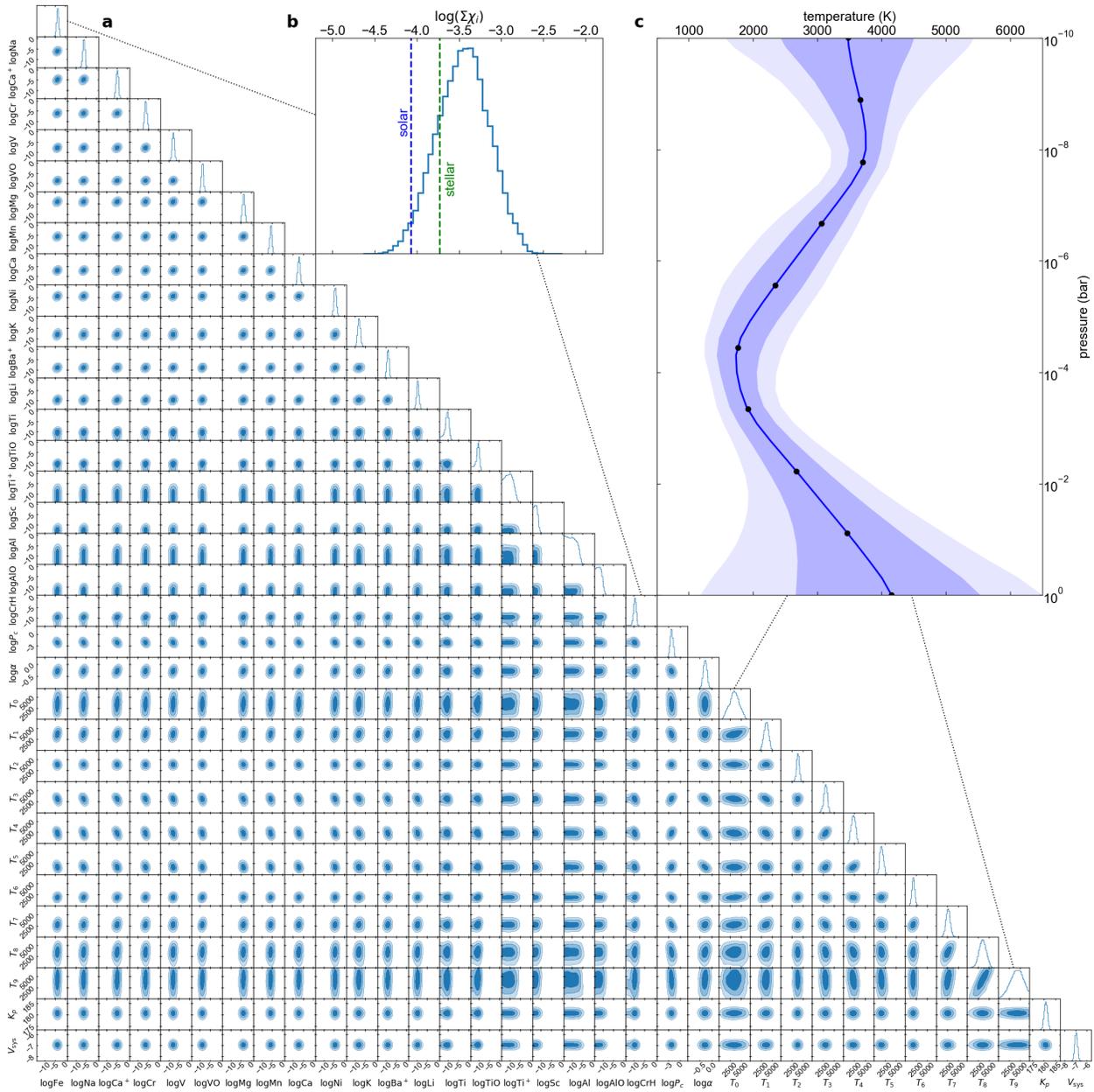

**Extended Data Figure 5 | Retrieved constraints on the atmospheric and orbital properties of WASP-76b obtained from three MAROON-X transits. a**, Corner plot of the marginalized posterior distributions for the abundance of included species, cloud-top pressure $P_c$ (in bars), scaling parameter α, temperature of different atmospheric layers, Keplerian velocity $K_p$, and systemic velocity $V_{sys}$. The shaded regions respectively depict the 39.3%, 86.5%, and 98.9% confidence intervals. **b**, The sum of the volume mixing ratio of individual metals, ions, and molecules included in the model. The equivalent sums for solar and stellar compositions (dashed lines) are also shown for comparison, with WASP-76b being consistent with slightly (+0.28 dex) super-stellar. **c**, The resulting vertical temperature structure from the ten temperature points ($T_0$ – $T_9$, black dots), showing the presence of a stratosphere.

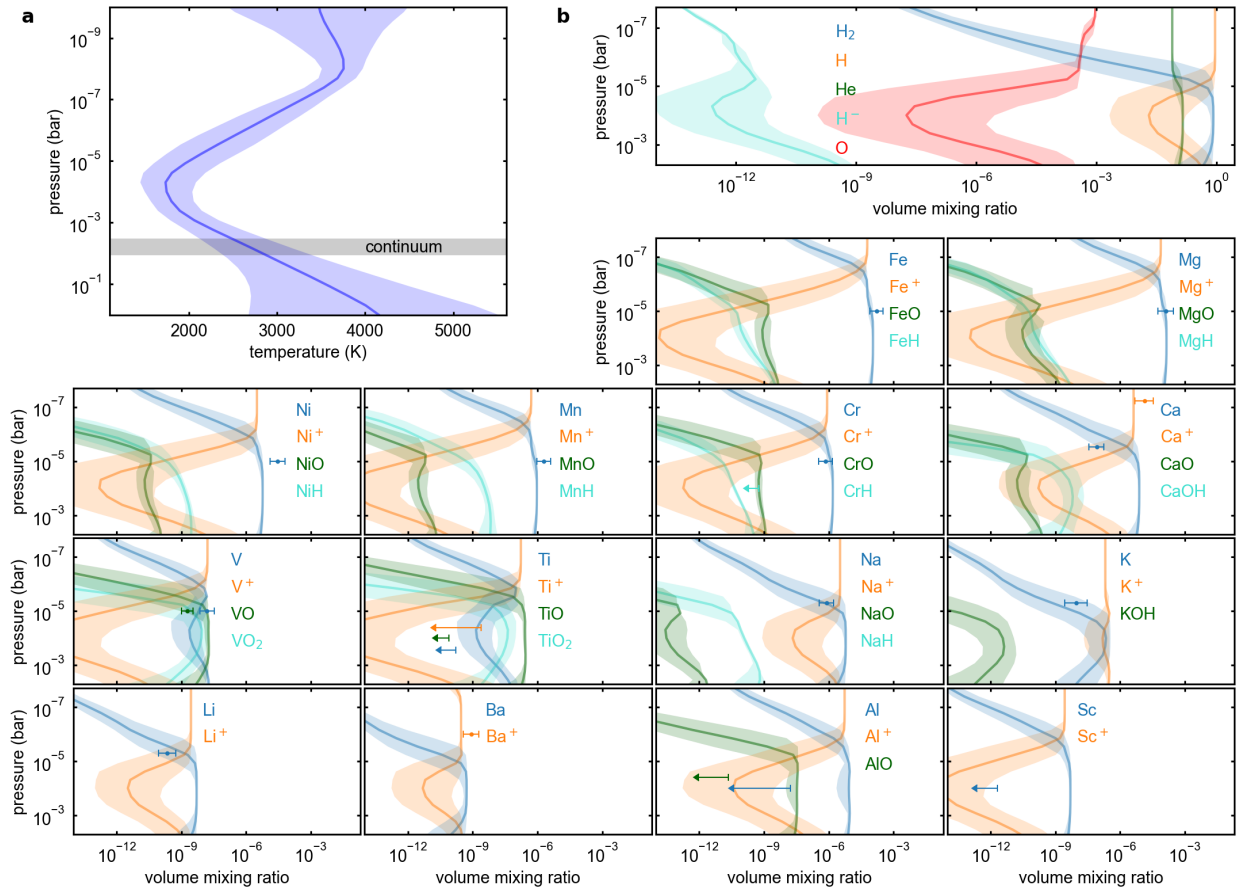

**Extended Data Figure 6 | Chemical equilibrium predictions of WASP-76b's atmospheric composition**. **a**, The retrieved temperature-pressure profile and cloud-top continuum pressure. **b**, Chemical equilibrium abundance predictions[58] for a wide range of elements given the retrieved temperature-pressure structure (panel a) and assuming a stellar atmospheric composition[18]. Measured abundances for WASP-76b at the estimated probed altitudes are also shown for comparison. Most refractory species (e.g., Fe, Mg, Ni, Mn, and Cr) are not expected to be significantly ionized below the microbar level and are relatively well-approximated by a constant-with-altitude volume mixing ratio model. With the exception of V and Ti, most elements are only expected to be bound in molecular form in trace amounts. Alkali metals, calcium, and barium all ionize more readily and have deep spectral features, and thus are expected to be significantly ionized at the lower pressures probed. Errorbars and shaded regions represent 1σ uncertainties.

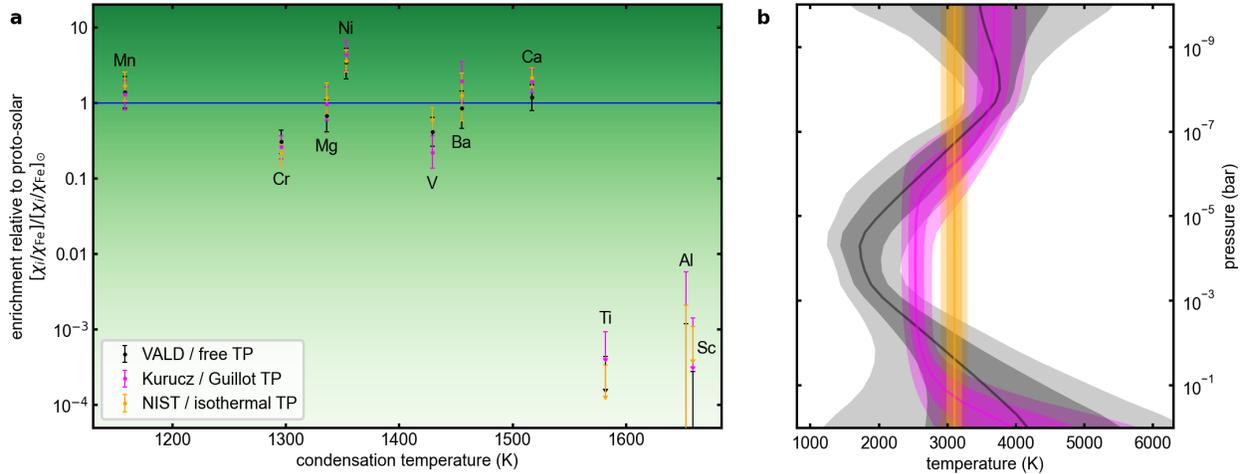

**Extended Data Figure 7 | Comparison of results from different retrieval prescriptions. a**, Retrieved abundance ratios given different model parameterizations. **b**, Inferred temperature profile for each associated retrieval prescription. The three retrievals use separate combinations of the VALD[52], Kurucz[53], and NIST[54] opacities, and fitted free[17], Guillot[70] and isothermal temperature-pressure profiles. Despite using differing temperature structure prescriptions and opacities, the recovered abundance ratios in all three retrievals are consistent within uncertainties. However, assuming an isothermal or Guillot profile can over-constrain the temperature structure. Errorbars represent 1σ uncertainties. Shaded regions represent 1σ and 2σ contours.

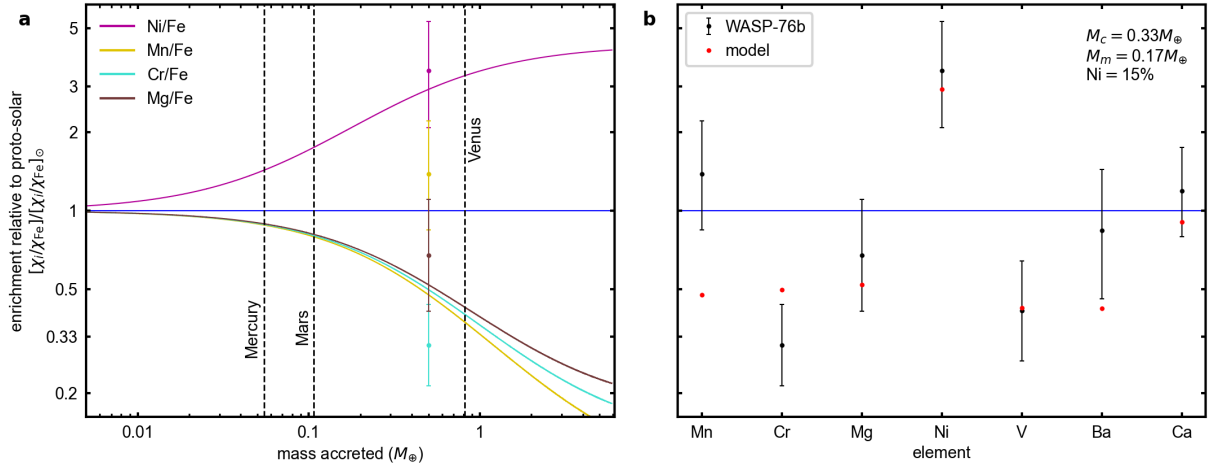

**Extended Data Figure 8 | Accretion toy model exploring the scenario of WASP-76b accreting a body with a Mercury-like composition. a**, The change in enrichment of elemental abundance ratios relative to proto-solar as a function of the accreted mass (V/Fe and Ba/Fe behave similarly to Mn/Fe and are not shown for clarity). In this example, the accreted body has a core-mass fraction of 1.95, a mantle composition matching that of Mercury's surface[84] and a core composition of 15% Ni and predominantly Fe as the rest[83]. The horizontal blue line denotes a proto-solar composition and the vertical dashed lines show the masses of Mercury, Mars, and Venus for reference. If too small a body is accreted ($\leq 0.1\,M_\oplus$) onto the initial $284\,M_\oplus$ of WASP-76b, the composition does not change significantly from proto-solar. If too big a body is accreted ($\geq 3\,M_\oplus$) the overall composition begins to change too drastically. While all abundance ratios require different masses to be perfectly matched under this assumed enrichment material composition, the overall best fit occurs if a large object between roughly the size of Mars and Venus is added to WASP-76b. **b**, A comparison between the data and the toy model assuming an accreted mass of $0.5\,M_\oplus$, where the data points except Mn/Fe are reasonably well matched. All errorbars denote 1σ uncertainties.

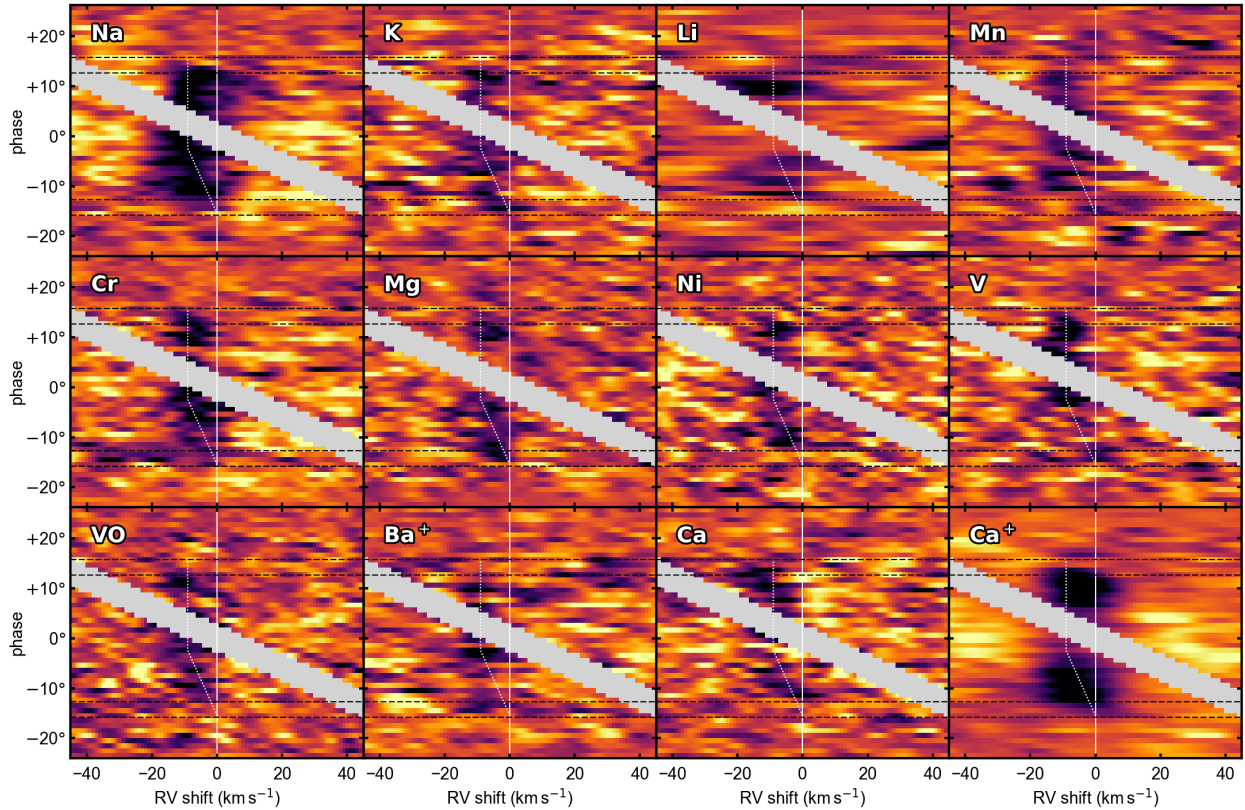

**Extended Data Figure 9 | Rest frame absorption signals of individual species on WASP-76b.** Shown in each panel are the cross-correlation trails for the species combined in Fig. 3b. Despite their wide range in condensation temperature, most species have a similar "kinked" absorption trail as Fe (white dotted line), likely indicating that condensation is not the sole culprit for the asymmetric signature. One notable exception is ionized calcium, which does not show such an asymmetry, likely due to Ca$^+$ triplet having an absorption depth consistent with probing an escaping atmosphere[18].

# Extended Data Table 1 | Retrieved abundance constraints for species on WASP-76b

| Element | $\log(\chi_i)$ | $\log(\chi_i/\chi_{\text{Fe}})$ | $\log(\frac{[\chi_i/\chi_{\text{Fe}}]}{[\chi_i/\chi_{\text{Fe}}]_\odot})$ | $\log(\frac{[\chi_i/\chi_{\text{Fe}}]}{[\chi_i/\chi_{\text{Fe}}]_*})$ | $T_{\text{cond}}$[20] [K] |
|---|---|---|---|---|---|
| Na | $-6.10^{+0.33}_{-0.34}$ | $-2.31^{+0.16}_{-0.17}$ | $-1.06^{+0.17}_{-0.18}$ | $-1.20^{+0.18}_{-0.19}$ | 958 |
| K | $-8.04^{+0.51}_{-0.52}$ | $-4.25^{+0.45}_{-0.43}$ | $-1.85^{+0.45}_{-0.43}$ | $-1.90^{+0.46}_{-0.43}$ | 1006 |
| Li | $-9.65^{+0.38}_{-0.41}$ | $-5.85^{+0.27}_{-0.31}$ | $-1.65^{+0.27}_{-0.32}$ | − | 1142 |
| Mn | $-5.70^{+0.33}_{-0.33}$ | $-1.91^{+0.19}_{-0.20}$ | $0.14^{+0.20}_{-0.21}$ | $0.16^{+0.21}_{-0.22}$ | 1158 |
| Cr | $-6.15^{+0.32}_{-0.32}$ | $-2.35^{+0.14}_{-0.15}$ | $-0.51^{+0.16}_{-0.16}$ | $-0.55^{+0.18}_{-0.18}$ | 1296 |
| CrH | $< -9.24$ | $< -5.52$ | $< -3.68$ | $< -3.71$ | 1296 |
| Fe | $-3.79^{+0.29}_{-0.31}$ | 0 | 0 | 0 | 1334 |
| Mg | $-3.87^{+0.35}_{-0.36}$ | $-0.08^{+0.21}_{-0.21}$ | $-0.17^{+0.21}_{-0.21}$ | $-0.15^{+0.22}_{-0.22}$ | 1336 |
| Ni | $-4.54^{+0.35}_{-0.35}$ | $-0.73^{+0.18}_{-0.21}$ | $0.54^{+0.19}_{-0.22}$ | $0.53^{+0.20}_{-0.23}$ | 1353 |
| V | $-7.80^{+0.34}_{-0.34}$ | $-4.00^{+0.18}_{-0.18}$ | $-0.44^{+0.20}_{-0.21}$ | − | 1429 |
| VO | $-8.72^{+0.26}_{-0.26}$ | $-4.94^{+0.19}_{-0.17}$ | $-1.37^{+0.21}_{-0.19}$ | − | 1429 |
| Ba | $-9.07^{+0.34}_{-0.36}$ | $-5.27^{+0.22}_{-0.26}$ | $-0.07^{+0.23}_{-0.26}$ | − | 1455 |
| Ca | $-7.07^{+0.33}_{-0.37}$ | $-3.28^{+0.15}_{-0.17}$ | $-2.11^{+0.16}_{-0.18}$ | $-2.18^{+0.19}_{-0.21}$ | 1517 |
| Ca$^+$ | $-4.87^{+0.40}_{-0.44}$ | $-1.09^{+0.16}_{-0.17}$ | $0.07^{+0.17}_{-0.18}$ | $-0.00^{+0.20}_{-0.21}$ | 1517 |
| Ti | $< -9.78$ | $< -5.97$ | $< -3.47$ | $< -3.48$ | 1582 |
| TiO | $< -10.10$ | $< -6.32$ | $< -3.83$ | $< -3.83$ | 1582 |
| Ti$^+$ | $< -8.61$ | $< -4.79$ | $< -2.30$ | $< -2.31$ | 1582 |
| Al | $< -7.77$ | $< -3.96$ | $< -2.92$ | − | 1653 |
| AlO | $< -10.67$ | $< -6.88$ | $< -5.84$ | − | 1653 |
| Sc | $< -11.67$ | $< -7.88$ | $< -3.56$ | − | 1659 |

⊙ Proto-solar

∗ Stellar